\documentclass{article}
\usepackage{spconf,amsmath,graphicx}
\usepackage{longtable}
\usepackage{subcaption}
\usepackage{multirow}
\usepackage{adjustbox}
\usepackage{graphicx}
\usepackage[table,xcdraw]{xcolor}
\usepackage{pgfplots}
\pgfplotsset{compat=1.14}

\usepackage{threeparttable}
\usepackage{booktabs}
\usepackage{verbatim}
\usepackage{geometry}
\title{MiTAS: A Compressed Time-domain Audio Separation Network with Parameter Sharing }

\name{Chao-I Tuan$^1$$^\star$ \qquad Yuan-Kuei Wu$^1$$^\star$ \qquad Hung-yi Lee$^1$ \qquad Yu Tsao$^2$ \thanks{$^\star$The two first authors made equal contributions} }
\address{
$^{1}$Graduate Institute of Communication Engineering, National Taiwan University\\
$^2$Research Center for Information Technology Innovation, Academia Sinica\\
\{chaoi111.t, ywk991112, tlkagkb93901106\}@gmail.com,
yu.tsao@citi.sinica.edu.tw
}

\begin{document}
%
\maketitle
\begin{abstract}
Deep learning methods have brought substantial advancements in speech separation (SS). 
Nevertheless, it remains challenging to deploy deep-learning-based models on edge devices. Thus, identifying an effective way to compress these large models without hurting SS performance has become an important research topic. Recently, TasNet and Conv-TasNet have been proposed. They achieved state-of-the-art results on several standardized SS tasks. Moreover, their low latency natures make them definitely suitable for real-time on-device applications. In this study, we propose two parameter-sharing schemes to lower the memory consumption on TasNet and Conv-TasNet. Accordingly, we derive a novel so-called MiTAS (Mini TasNet). Our experimental results first confirmed the robustness of our MiTAS on two types of perturbations in mixed audio. We also designed a series of ablation experiments to analyze the relation between SS performance and the amount of parameters in the model. The results show that MiTAS is able to reduce the model size by a factor of four while maintaining comparable SS performance with improved stability as compared to TasNet and Conv-TasNet. 
This suggests that MiTAS is more suitable for real-time low latency applications.

\end{abstract}

\begin{figure*}[t!]
    \centering
    \includegraphics[scale=0.09]{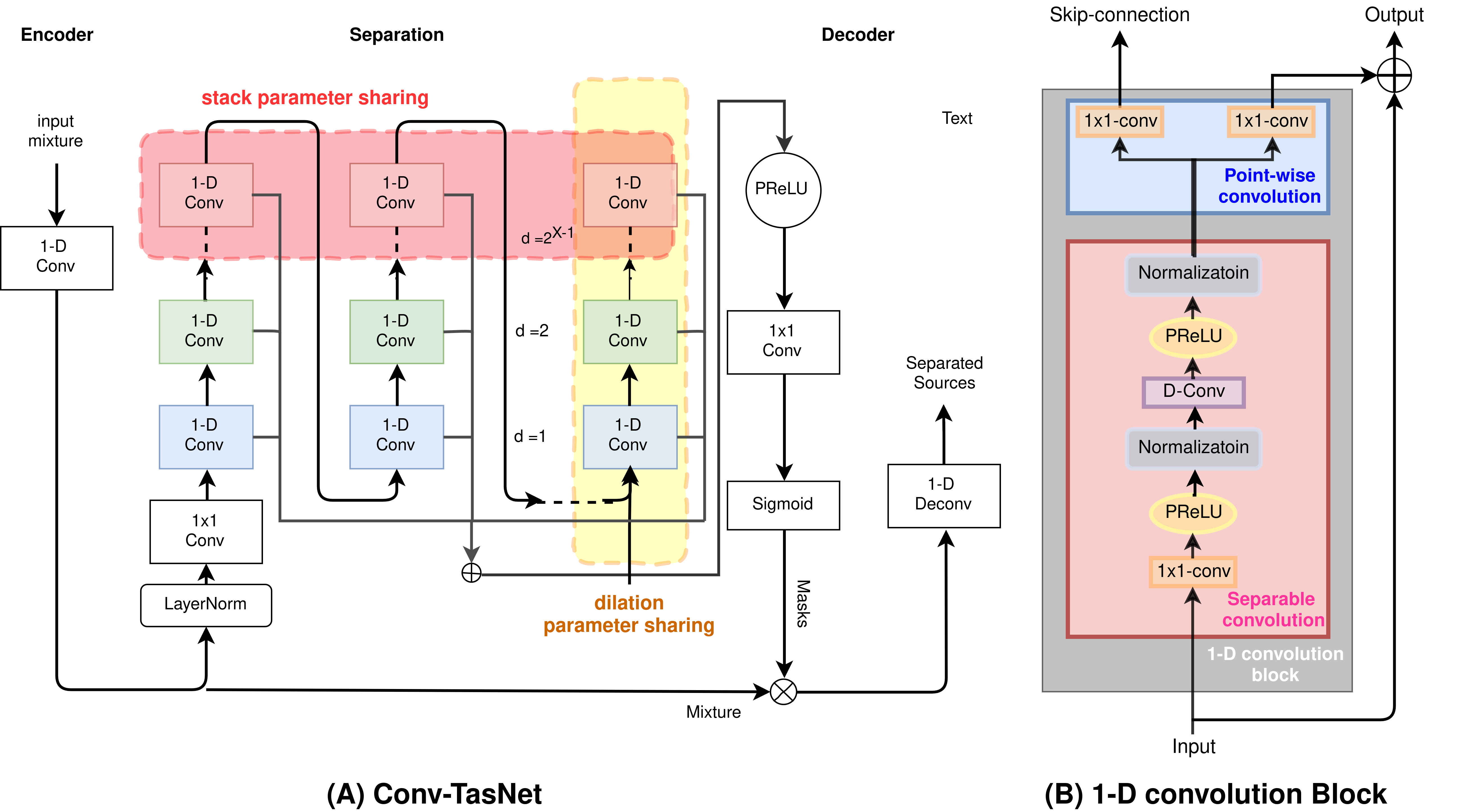}
    \captionsetup{font=small}
    \caption{(A)The whole model of Conv-TasNet. There are slightly different between the architectures of TasNet and Conv-TasNet, but the concepts of sharing mechanism are the same. Stack parameter sharing is colored with red, and dilation parameter sharing is colored with yellow. The selected base component shares the same weight in the colored area. (B)An 1-D convolutional block consists of two base components. One is separable convolution and the other is pointwise convolution.}
    \label{fig:model}
\end{figure*}

\section{Introduction}
\label{sec:intro}
While being in a crowded or noisy environment, such as a party or a sports game, a combination of ambient noises and interferences can distract our attention. Nevertheless, we are still able to focus on the voice produced by a specific speaker. The phenomenon underlying our brain's ability to concentrate on a particular voice while filtering out background noise is called the cocktail party effect. However, it is still a challenging task for computers to completely solve the cocktail party problem, which is also known as the speech separation (SS) task. 
When occurring in real-world scenarios, a speech communication often consists of multiple speakers. Thus, it is favorable to have a speaker voice separated from overlapping speeches  to successfully complete the downstream tasks, such as speech recognition, speaker 
identification, and audio classification. 

A great effort has been made to tackle the SS problem, and deep learning techniques have enabled notable progresses in recent years \cite{Supervised, dnntraining, deep_cluster, deep_attractor, pit, tasnetv1,Gender, RNN}. 
Among deep-learning-based SS approaches, deep clustering (DC) \cite{deep_cluster,deep_cluster_single} and permutation invariant training (PIT) \cite{pit, pit_multitalker}
stand out. Both of them use spectrograms as acoustic features, which are obtained by applying the short-time Fourier transform (STFT) on speech waveforms.

It has been reported that T-F masking methods can provide satisfactory performance in many SS tasks. However, waveform reconstruction cannot be perfect owing to the inaccurate phase information.  Furthermore, a long temporal window in STFT is needed owing to the the requirements of high-resolution frequency decomposition. The increased latency results in less applicability on real-time devices.
To overcome the above limitations, time-domain SS has recently drawn much attention. A notable example is the TasNet model and its extensions \cite{lstm_tasnet, tasnetv1, tasnetv3}
, which have achieved state-of-the-art performance on WSJ0-2mix\cite{2017wsj0, deep_cluster_single} (a standardized SS 
dataset). Waveform-domain SS methods alleviate the disadvantages of sub-optimal feature representations (owing to decoupling of phase and magnitude of the signal) and long latency in calculating the spectrograms.

However, another limitation further confines the applicability of time-domain SS for on-device applications: a deep-learning-based model usually has a large model size and thus requires large memory storage. For example, uPIT-BLSTM-ST has nearly 92.7M parameters \cite{pit_multitalker}, 
Chimera++ has 32.9M parameters \cite{chimera}, LSTM-TasNet has 32.0 M parameters \cite{lstm_tasnet}, 
DPCL++ has 13.6 M parameters \cite{deep_cluster_single}; the previous version of TasNet contains 8.9M 
\cite{tasnetv1}, and the latest Conv-TasNet still has approximately 5.1M parameters \cite{tasnetv3}.

Significant research interests have arisen concerning the exploration of model compression to meet the desire for more well-suited models for edge and mobile devices \cite{mobilenet, factorized_cnn}. Popular model compression techniques include knowledge distillation \cite{distill}, parameter pruning and vector quantization \cite{ppvq}. Based on the success of ALBERT \cite{albert} in natural language processing, it demonstrates a proof of concept for large network can be directly trained on parameters sharing and factorize representation, leading to significantly parameter downscaling without affecting performance.

In this paper, we extended a previous work on a fully-convolutional time-domain audio separation network (Conv-TasNet) \cite{tasnetv1,tasnetv3} to further solve its expensive storage problem by analyzing the feasibility of parameter reduction based on its model architecture. Conv-TasNet is formed by a 1-D convolutional encoder and decoder along with a temporal convolution network (TCN-)based separation module. Among these three parts, the separation module constitutes the major portion of the entire model size. Thus, we decided to aim at the separation module and derive a novel scheme for parameter-sharing to reduce its complexity and also increase the overall training speed of TasNet. As a result, we derive the so-called Mini TASNET (MiTAS) architecture, which has comparable or even better performance while reducing by nearly 50\% the number of parameters as compared to the latest version of Conv-TasNet. Moreover, experimental results show that the performance robustness against background noise can be improved through the proposed parameter-sharing scheme. 


\section{The MiTAS model}
\label{sec:The MiTAS model}

\subsection{Architecture}
The architecture of MiTAS is based on TasNet and Conv-TasNet and consists of an encoder, a separation module, and a decoder. The encoder first maps a segment of mixed waveforms into a high-dimensional representation; then, the separation model estimates a mask to generate the masked and separated features; finally, the decoder reconstructs the  separated source waveform. 
TasNet and Conv-TasNet\cite{tasnetv1,tasnetv3} use a fully convolutional separation module, which consists of a stack of 1-D dilated convolution blocks. The dilation of each block is increased by an order of 2 up to a given threshold, and the whole stack structure is repeated several times. Such a dilated convolutional structure has also been used as a fundamental block in Wavenet \cite{wavenet}. This enables models with a larger receptive field and fewer layers.

The separation module constitutes the major portion of the overall model size. In both TasNet and Conv-TasNet models, the TCN separation module 
contributes to nearly 98\% of the total number of parameters. In Figure \ref{fig:model}, we illustrate the components in the 1-D convolution block, which is the backbone of the separation module.
A 1-D convolution block in a time-dilated convolutional network is composed of a separable convolution and a point-wise convolution operation.  

\begin{figure}[b!]
    \centering
    \includegraphics[scale=0.15]{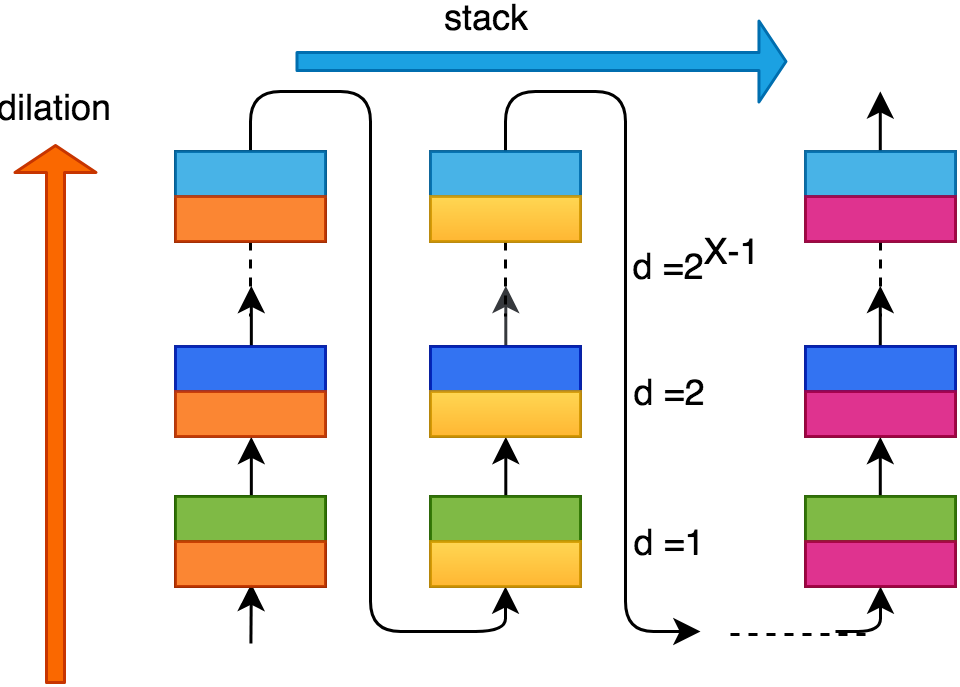}
    \captionsetup{font={small},singlelinecheck=off,justification=raggedright}
    \caption{Example of "share through dilations" for separable conv. and "share through stacks" for point-wise conv.}
    \label{fig:example}
\end{figure}

\begin{table*}
\centering
\begin{threeparttable}
\caption{The effect of different parameter sharing configurations in MiTAS.} 
\label{tab:exp}
\begin{tabular}{lllllllll}
\hline
\textbf{Ablation Model\tnote{*}}    & \multicolumn{2}{l}{\textbf{\begin{tabular}[c]{@{}l@{}}separable\\ conv.\end{tabular}}} & \multicolumn{2}{l}{\begin{tabular}[c]{@{}l@{}}point-wise\\     conv.\end{tabular}} & \textbf{Size}                        & \textbf{C.P.}                          & \textbf{\begin{tabular}[c]{@{}l@{}}SI-SNRi \\    (dB)\end{tabular}} & \textbf{\begin{tabular}[c]{@{}l@{}}SDRi\\  (dB)\end{tabular}} \\
( shared dimension)                                                                                    & stack                                       & dil.                                       & stack                                    & dil                                     &                                      &                                        &                                                                     &                                                               \\ \hline
\rowcolor[HTML]{EFEFEF} 
TasNet (Base-Model)                    & \textbf{-}                                   & \textbf{-}                                & \textbf{-}                               & \textbf{-}                              & \textbf{8.9 M}                       & 100\%                                  & \textbf{16.15}                                                      & \textbf{16.41}                                                \\
MiTAS\_ns \tnote{**}                          & -                                            & -                                         & V                                        & -                                       & 5.6M                                 & 64\%                                   & 15.61                                                               & 15.88                                                         \\
MiTAS\_sn                           & V                                            & -                                         & -                                        & -                                       & 5.5M                                 & 62.7\%                                 & 15.59                                                               & 15.86                                                         \\
MiTAS\_nd                           & -                                            & -                                         & -                                        & V                                       & 5.1M                                 & 58.7\%                                 & 15.28                                                               & 15.55                                                         \\
MiTAS\_dn                           & -                                            & V                                         & -                                        & -                                       & 4.9M                                 & 56.51\%                                & 15.08                                                               & 15.35                                                         \\
MiTAS\_na                           & -                                            & -                                         & V                                        & V                                       & 4.7M                                 & 53.5\%                                 & 15.03                                                               & 15.31                                                         \\
\textbf{MiTAS\_ss}                  & \textbf{V}                                   & \textbf{-}                                & \textbf{V}                               & \textbf{-}                              & {\color[HTML]{FE0000} \textbf{2.3M}} & {\color[HTML]{FE0000} \textbf{26.7\%}} & {\color[HTML]{FE0000} \textbf{15.01}}                               & {\color[HTML]{FE0000} \textbf{15.42}}                         \\
\textbf{     ---- with H=1024}              & \textbf{V}                                   & \textbf{-}                                & \textbf{V}                               & \textbf{-}                              & {\color[HTML]{000000} \textbf{4.5M}} & {\color[HTML]{000000} \textbf{52.2\%}} & {\color[HTML]{000000} \textbf{15.30}}                               & {\color[HTML]{000000} \textbf{15.56}}                         \\
\textbf{MiTAS\_ss (6 stacks)}       & \textbf{V}                                   & \textbf{-}                                & \textbf{V}                               & \textbf{-}                              & {\color[HTML]{FE0000} \textbf{2.3M}} & {\color[HTML]{FE0000} \textbf{26.7\%}} & {\color[HTML]{FE0000} \textbf{14.84}}                               & {\color[HTML]{FE0000} \textbf{15.15}}                         \\ \hline
Simplified-Base-Model 1           &\multicolumn{4}{l}{only 1 stack}          & 2.3M                                 & 26.7\%                                 & 13.10                                                               & 13.29                                                         \\ 
Simplified-Base-Model 2                & \multicolumn{4}{l}{only 1 block in each stack}          & 1.3M                                 & 14.6\%                                 & 8.53                                                               &   8.76                                                       \\ \hline
\rowcolor[HTML]{EFEFEF} 
Conv-TasNet (Base-Model)               & -                                            & -                                         & -                                        & -                                       & \textbf{5.1M}                        & \textbf{100\%}                         & \textbf{14.79}                                                      & \textbf{15.05}                                                \\
Conv-MiTAS\_sn                     & V                                            & -                                         & -                                        & -                                       & 3.9M                                 & 77\%                                   & 14.28                                                               & 14.55                                                         \\
Conv-MiTAS\_nd                     & -                                            & -                                         & V                                        & -                                       & 2.9M                                 & 58\%                                   & 14.28                                                               & 14.55                                                         \\
\textbf{Conv-MiTAS\_ss}            & \textbf{V}                                   & \textbf{-}                                & \textbf{V}                               & \textbf{-}                              & {\color[HTML]{FE0000} \textbf{1.8M}} & {\color[HTML]{FE0000} \textbf{36\%}}   & {\color[HTML]{FE0000} \textbf{14.22}}                               & {\color[HTML]{FE0000} \textbf{14.49}}                         \\
\textbf{     ---- with H=1024}              & \textbf{V}                                   & \textbf{-}                                & \textbf{V}                               & \textbf{-}                              & \textbf{3.4M}                        & \textbf{68\%}                          & \textbf{14.93}                                                     & \textbf{15.21}                                                     \\
\textbf{Conv-MiTAS\_ss (5 stacks)} & \textbf{V}                                   & \textbf{-}                                & \textbf{V}                               & \textbf{-}                              & {\color[HTML]{FE0000} \textbf{1.8M}} & {\color[HTML]{FE0000} \textbf{36\%}}   & {\color[HTML]{FE0000} \textbf{14.12}}                               & {\color[HTML]{FE0000} \textbf{14.42}}                         \\ \hline
Simplified-Base-Model 1                 &\multicolumn{4}{l}{only 1 stack}      & 1.8M                                 & 36\%                                   & 12.50                                                               & 12.76                                                         \\ 
Simplified-Base-Model 2               & \multicolumn{4}{l}{only 1 block in each stack}               & 0.8M                                 & 16\%                                   & 8.01                                                               & 8.27                                                         \\ \hline
\end{tabular}
\begin{tablenotes}
\footnotesize
\item[*] Only partial results are listed in the table. The whole results can be found in Fig 3
\item[**] For each separable, point-wise conv., the abbreviations stand for: \textit{'s'}: shared in stack dimension, \textit{'d'}: shared in dilation dimension, \textit{'a'}: share in both dimension, \textit{'n'}: no sharing operation
 \end{tablenotes}
\end{threeparttable}
\end{table*}

\subsection{Cross-Layer parameter sharing}
To compress the TasNet-based model, we propose using the cross-layer parameter 
sharing to optimize the parameter efficiency. In particular, in this study we designed the sharing operation that will be later conducted on the base essential components of a 1-D convolutional block, including depth-wise separable convolution and point-wise convolution. The 1-D convolutional block is illustruated in Figure \ref{fig:model}. First, we selected the base components to share, either for the depth-wise separable convolution, the point-wise convolution, or both(i.e. the whole 1-D convolutional block). As shown in Figure \ref{fig:model}, the two weight sharing schemes we propose are horizontal (in stack dimension) and vertical (in dilation dimension), respectively. The 
expression "share through stacks" means that all stacks are shared with the same parameters. In other words, all base components with the same dilation share the same weights. By contrast, the expression "share through dilations" means that all base components with different dilations in the same stack share the same weights. Take Figure \ref{fig:example} as an example. The separable convolutions are shared through dilations while the point-wise convolutions are shared through stacks.

\section{Experiments}
\label{sec:Experiments}
\subsection{Datasets}
We evaluated our system on the WSJ0-2mix two-speaker SS task \cite{2017wsj0, deep_cluster_single}. This task contains 30 hours of training data and 
10 hours of validation data generated from the WSJ0 si\_tr\_s dataset. The speech utterances were then mixed up at various signal-to-noise ratio (SNR) levels between -5 dB to 5 dB.
The 5-hour testing set was similarly generated from the WSJ0 si\_dt\_05 and 
waveforms were downsampled to 8KHz as a preprocessing step.

The MUSAN dataset was additionally used to test the SS performance robustness against background noise. We only used the noise data of this dataset, which contains 6 hours of noise samples that can be divided into two categories. One is technical noises, such as dialtones and fax machine, whereas the other category is ambient sounds, such as thunders, footsteps and animal noises. The noise signals were resampled to 8 KHz. Then, the mixed signals were generated by adding a randomly selected noise sample to the test utterance at a specific SNR level.

\subsection{Ablation study}

To find the best weight sharing scheme for MiTAS, we conducted ablation studies on all possible candidate models, including 15 variants for both TasNet and Conv-TasNet.
We considered {\it separable conv.} and {\it point-wise conv.} in the 1-D conv. block as the base components. Thus, we could decide whether to perform parameter sharing in stack, dilation ,or both dimensions. As a result, we obtained 16 sets of results, including the 15 model variants and the original TasNet and Conv-TasNet models (without any parameter sharing). We denoted the parameter-sharing schemes by subscripts. For example, MiTAS\_ss is the abbreviation for TasNet with {\it separable conv.} and {\it point-wise conv.} parameters shared in the stack dimension, respectively; likewise, Conv-MiTAS\_ad is the abbreviation for Conv-TasNet with {\it separable conv.} shared in all dimensions and {\it point-wise conv.} shared in the dilation dimension.

\subsection{Results}
In Table 1, we list some results from the 16 models and the original TasNet and Conv-TasNet models, denoted as TasNet (Base-Model) and Conv-TasNet (Base-Model), respectively. The evaluation is based on two standardized metrics, namely source-to-distortion ratio improvement (SDRi) \cite{sdr} and the scale-invariant source-to-noise ratio improvement (SI-SNRi) \cite{siSNR}.
In this experiment, we intended to investigate the relation between different parameter-sharing schemes and the achievable SS performance. 
The table is sorted by model size; the compression ratios with respect to the Base-Models are also listed. Note from this table that MiTAS with all types of parameter sharing schemes can deliver comparable or only slightly decreased SS performance as compared to the two Base-Models. 
 


To further investigate whether parameter-sharing schemes in both dimensions indeed improve the parameter efficiency, we implemented two other simplified models (denoted as Simplified-Base-Model 1 and Simplified-Base-Model 2). 
Simplified-Base-Model 1 contains a single stack with the architecture of the base model and has the same number of parameters as the stack-sharing ablation models (MiTAS\_ss and Conv-MiTAS\_ss). Simplified-Base-Model 2 contains one single block in each stack in the base model and has the same number of parameters as the dilation sharing ablation models( MiTAS\_dd and Conv-MiTAS\_dd). From Table 1, we observe notable performance drops for the simplified models. 

For both MiTAS and Conv-MiTAS variants, we noticed consistent trends in which sharing parameters among each stacks better preserves SS performance while still achieving the maximum compression, i.e., 26.7\% and 36\% of the size of the original Base-Models, respectively. Given that MiTAS\_ss has a better SS performance than Conv-MiTAS\_ss, MiTAS\_ss is chosen as a representative model in the following experiments.  

We also increased the number of stacks under stack-sharing condition. The performance is slightly improved while the model size remains unchanged. Moreover, we tried to enlarge the dimensions of both separable conv. and point-wise conv.; however, we did not see a large gain on performance because of the trade-off with the model size. 

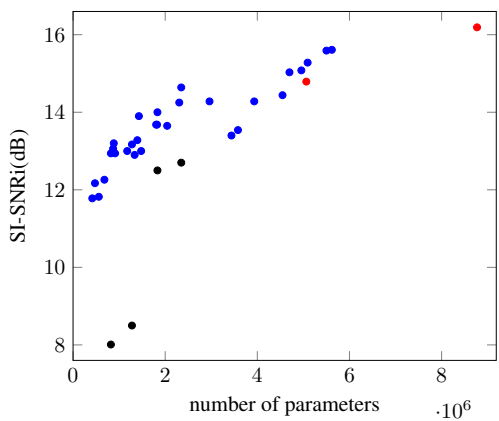
\begin{figure}[b!]
\centering
\begin{minipage}{\linewidth}
\centering
\resizebox{.8\linewidth}{!}{
\begin{tikzpicture}
\begin{axis}[
    enlargelimits=0.05,
    xlabel=number of parameters,
    ylabel=SI-SNRi(dB)
]
\addplot+ [
        only marks,
        red,mark options={
		fill=red,scale=0.8},mark=*,
    ] table[meta=model]{tas.dat};
\addplot+ [
        only marks,
        blue,mark options={
		fill=blue,scale=0.8},mark=*,
    ] table[meta=model]{al.dat};
\addplot+ [
        only marks,
        black,mark options={
		fill=black,scale=0.8},mark=*,
    ] table[meta=model]{opt.dat};
\end{axis}
\end{tikzpicture}
}
\captionsetup{font={small},singlelinecheck=off,justification=raggedright}
\caption{SI-SNRi versus number of parameters for all models.
Red dots denote the origin models(TasNet and Conv-TasNet). Black plots denote the baseline models, which reduce the number of stacks or 1-D convolutional blocks in each stack. All of our parameter-sharing models are colored in blue.}

\label{fig:plot}
\end{minipage}

\end{figure}

\subsection{Effect of Parameter Reduction}
In this section, we intended to further investigate the correlation between the amount of parameters and the SS results. Figure 3 shows the SI-SNRi results versus the number of parameters for MiTAS, TasNet and Conv-TasNet. Note from this figure that a clear performance drop occurs as the amount of parameter reduces. 
Notably, we can see that while in the same degree of parameter number, all variants of MiTAS perform better than both base and simplified base models. There are two simplified base models whose SI-SNRi drops significantly with number of parameters reduced below 2 M. In contrast, our parameter-sharing scheme still achieves a relatively high performance with such parameter reduction scale.

\subsection{Robustness and sensitivity of the proposed models}
As mentioned earlier, the objective of this study is to investigate an effective way to compress deep-learning-based models so that a SS system can be deployed in resource-constrained devices, such as mobile and edge devices. In addition to reducing model sizes, performance robustness against background noise is another consideration. We tested the proposed MiTAS under two circumstances: with background noise and with starting-point shifting.



\subsubsection{Sensitivity to noise}
A robust model should be insensitive to background noise. That is, the performance difference between noisy and quiet environments should be as small as possible. We evaluated the MiTAS SS performance under noise conditions, where mixed speech signals were further contaminated with two types of noises, namely Gaussian noise and real word noise(MUSAN dataset) at 0dB, 3dB and 5dB signal-to-noise ratio (SNR). We used the MiTAS\_ss model and its enlarge channel-size version (where we swept H from 512 to 1024) in this experiment.
The two models have better SS performance than both TasNet and Conv-TasNet Base model and are more parameter efficient. The results are presented in Table 2. Note from this table that the performance of MiTAS\_ss (and also MiTAS\_ss-(H=1024)) is better than TasNet in almost every conditions. Considering this, our parameter-sharing scheme is more robust against noisy environments.

\begin{table}[htb!]
\captionsetup{font={small}}
\caption{SI-SNRi scores under noisy situation. We prepared noisy testing data of 0, 3 and 5 dB SNR levels. Both MiTAS\_ss and MiTAS\_ss-(H=1024) outperform TasNet in almost all the situations.}
\begin{threeparttable}
\resizebox{.50\textwidth}{!}{
\begin{tabular}{cccccccc}
\cline{1-7}
                     & \multicolumn{2}{c}{0 dB} & \multicolumn{2}{c}{3 dB} & \multicolumn{2}{c}{5 dB} &  \\
                     & Gau\tnote{*}     & Mus\tnote{*}     & Gau     & Mus     & Gau     & Mus     &  \\ \cline{1-7}
TasNet               & 2.39         & 3.64      & 4.2          & 5.04      & 5.98         & 6.29      &  \\
MiTAS\_ss         & 2.55         & 3.79      & 4.57         & 5.11      & 5.98         & 6.12      &  \\
\ \ ---(H=1024) & 2.83         & 3.99      & 5.05         & 5.26      & 6.38         & 6.33      &  \\ \cline{1-7}
\end{tabular}
}
\label{table:noise}

\begin{tablenotes}
\footnotesize
\item[*] \text{Gau} and \text{Mus} demote the Gaussian and real world noise, respectively.
\end{tablenotes}
\end{threeparttable}

\end{table}

\subsubsection{Sensitivity to starting point shifting}
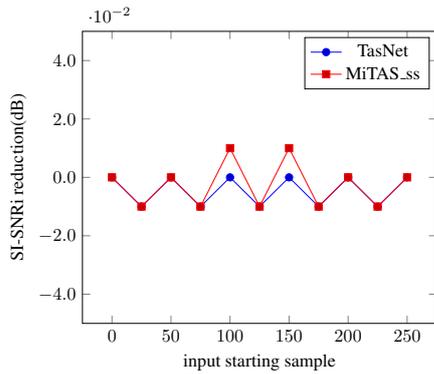
\begin{figure}[t!]
\centering
 \resizebox{.7\linewidth}{!}{
    \begin{tikzpicture}
    \begin{axis}[xlabel= input starting sample ,ylabel= SI-SNRi reduction(dB),
    enlarge y limits = 2,
    y tick label style={
        /pgf/number format/.cd,
            fixed,
            fixed zerofill,
            precision=1,
        /tikz/.cd
    },
    ]
    \addplot coordinates {
    (0,0) (25, -0.01) (50, 0)(75,-0.01) (100,0) (125,-0.01)(150, 0)(175,-0.01) (200,0) (225,-0.01) (250,0)
    };
    \addlegendentry{TasNet}
    \addplot coordinates {
    (0,0) (25, -0.01) (50, 0)(75,-0.01) (100,0.01) (125,-0.01)(150, 0.01)(175,-0.01) (200,0) (225,-0.01) (250,0)
    };
    \addlegendentry{MiTAS\_ss}
    \end{axis}
    \end{tikzpicture}
}
    \captionsetup{font={small}}
    \caption{The SI-SNRi reduction(compared to non-shifting case) caused by shifting the starting point from 25 to 250 samples with a common difference of 25.}
    \label{fig:shift}
\end{figure}

Different from natural language processing, where the word sequence has a deterministic starting point, a robust SS model should have the ability to separate a specific audio from any starting point of the mixed audio. Therefore, we reproduced the experiment mentioned in the TasNet \cite{tasnetv1} and Conv-Tasnet \cite{tasnetv3} papers. To evaluate the certain robustness, we chose different starting points ranging from 25 to 250 sample points. Then, we compared the separation performance of each model on the same example mixture with a given value of input shift. A shift of \textit{s} samples corresponds to starting the separation at \textit{s} instead of corresponding starting point. Figure \ref{fig:shift} shows that our model has the same robustness as TasNet despite the great reduction on parameter size.

\section{Conclusions}
In summary, MiTAS represents a significant step toward the realization of SS on edge devices. We investigated some of the important design aspects leading to a parameter-efficient model. Our MiTAS model is much smaller than Tasnet and Conv-Tasnet while yielding comparable SS performance and even better robustness against background noise. 
These advantages make MiTAS especially suitable for real-world low-resource devices. 
The proposed approach is complementary to other model compression techniques such as knowledge distillation and parameter quantization. Our future work will aim to integrate the proposed parameter-sharing approach with other compression techniques.
Moreover, we believe that existing performance-boosting strategies can also be combined with MiTAS to attain even better performance.

\bibliographystyle{IEEEbib}
\bibliography{strings,refs}

\end{document}